\documentclass{appolb}
\usepackage{epsfig}

\usepackage{amsmath}
\usepackage{bbm}
\usepackage{bm}
\usepackage{latexsym}
\usepackage{graphicx}
\usepackage{subfigure}% Include figure files
\usepackage{amsfonts}
\newcommand{\uni}[1]{\,\mathrm{#1}}
\newcommand{\beq}{\begin{equation}}
\newcommand{\eeq}{\end{equation}}

\begin{document}

\title{\bf Maximal Entropy Random Walk:\\ solvable cases of dynamics\thanks{Presented by J.K.O. at the 24th Marian Smoluchowski Symposium on Statistical Physics (Zakopane, Poland, September 17-22, 2011).}
}

\author{J.K. Ochab\footnote{jeremi.ochab@uj.edu.pl}
\address{Marian Smoluchowski Institute of Physics, Jagiellonian University,\\
 Reymonta 4, 30-059 Krak\'ow, Poland}}

\maketitle

\begin{abstract}
We focus on the study of dynamics of two kinds of random walk: generic random walk (GRW) and maximal entropy random walk (MERW) on two model networks: Cayley trees and ladder graphs. The stationary probability distribution for MERW is given by the squared components of the eigenvector associated with the largest eigenvalue $\lambda_0$ of the adjacency matrix of a graph, while the dynamics of the probability distribution approaching to the stationary state depends on the second largest eigenvalue $\lambda_1$.

Firstly, we give analytic solutions for Cayley trees with arbitrary branching number, root degree, and number of generations. We determine three regimes of a tree structure that result in different statics and dynamics of MERW, which are due to strongly, critically, and weakly branched roots. We show how the relaxation times, generically shorter for MERW than for GRW, scale with the graph size. 

Secondly, we give numerical results for ladder graphs with symmetric defects. MERW shows a clear exponential growth of the relaxation time with the size of defective regions, which indicates trapping of a particle within highly entropic intact region and its escaping that resembles quantum tunneling through a potential barrier. GRW shows standard diffusive dependence irrespective of the defects.
\end{abstract}

\PACS{05.40.Fb, 02.50.Ga, 89.75.Hc, 89.70.Cf;}

\section{Introduction}

After Einstein \cite{Einstein} and Smoluchowski \cite{Smoluch} gave explanations of Brownian motion and originated the theory of diffusive processes, there has been an unceasing research on models of random walk (RW), which may be regarded as time or space discretization of these processes. Thousands of papers and textbooks in statistical physics, particle physics, engineering, economics, biophysics, etc., have been and still are published.

From mathematical perspective, RW is a Markov chain describing the random consecutive steps of a particle. As an example, the well-known Polya random walk on a lattice \cite{Polya} at each time performs equiprobable steps to any of the neighboring nodes. This process, generalized to any graph, is known as the ordinary or generic random walk (GRW).

RW can also maximize the entropy of paths, and hence we call it the maximal entropy random walk (MERW); lately, this type has been studied in \cite{ZB1,ZB2}.
This principle of entropy maximization, which is a global one alike the least action principle, earlier brought about the biological concept of evolutionary entropy \cite{Evolutionary1,Evolutionary2}. 
It also served as an optimal sampling algorithm in the problem of importance sampling \cite{H}.
MERW has also begun to be used in the study of complex networks \cite{MERW+CN1,MERW+CN2,MERW+CN3,MERW+CN4,MERW+CN5}.

The defining feature of MERW makes the paths of given length and end-points equiprobable. This leads to an unprecedented feature that the stationary probability on diluted lattices localizes in the biggest spherical region \cite{ZB1,ZB2}. An interactive online demonstration \cite{BW} illustrates this feature. In this paper, we focus on how the dynamics of GRW and MERW differs. More precisely, we show analytic expressions for stationary probability distributions and relaxation times of GRW and MERW on Cayley trees; we also give numerical results for ladder graphs, showing that the relaxation time for MERW grows exponentially with the size of defective regions as opposed to diffusion behavior for GRW.

In this paper, in Sec. 2 we provide definitions and notes on the two types of random walk. In Sec. 3, we give several analytical results concerning Cayley trees (involving eigenproblem solution for the adjacency matrix, discussion of stationary state and relaxation). Lastly, in Sec. 4, we show numerical results concerning relaxation process on a class of ladder graphs.

\section{General considerations}

Let us consider a discrete time random walk defined by a constant stochastic matrix $\mathbf{P}$, on a finite connected undirected graph. 
The probability that a random walker which can be found on a node $i$ at time $t$ hops to a node $j$ at time $t+1$ is encoded by the element $P_{ij} \ge 0$ of this matrix.
Another condition fulfilled by these matrix element is $\sum_j P_{ij} = 1$ for all $i$.
If we denote by $\mathbf{A}$ the adjacency matrix of the graph ($A_{ij}=1$ if $i$ and $j$ are neighbors, and $A_{ij}=0$ otherwise),
we can formulate an additional condition: $P_{ij}\leq A_{ij}$, which means that particles are allowed to jump between neighboring nodes only.
The stochastic matrix corresponding to the generic random walk (GRW) is given by:
\beq
P_{ij} = \frac{A_{ij}}{k_i} \ ,
\label{Porw}
\eeq
where $k_i = \sum_j A_{ij}$ is the node degree, and the probability of selecting one of $k_i$ neighbors of the node $i$ is uniform.
This means that the entropy of neighbor selection is maximized and shows that this is the standard Einstein-Smoluchowski-Polya random walk.
Lastly, the stationary state of GRW is given by $\pi_i = k_i/\sum_j k_j$.

On the other hand, maximal entropy random walk (MERW) maximizes the entropy of choosing a trajectory of given length and end-points. 
This principle leads to
\beq
P_{ij} = \frac{A_{ij}}{\lambda_0} \frac{\psi_{0j}}{\psi_{0i}},
\label{Pmerw}
\eeq
where $\lambda_0$ is the largest eigenvalue of the adjacency matrix 
$\mathbf{A}$ and $\psi_{0i}$ is the $i$-th component of the corresponding eigenvector $\vec{\psi}_0$. From the Frobenius-Perron theorem and from the fact that the adjacency matrix $\mathbf{A}$ is irreducible it follows that all elements of $\vec{\psi}_0$ are strictly positive.
Shannon-Parry measure \cite{P} then describes the stationary state of $\mathbf{P}$:
\beq
\pi_i = \psi^2_{0i} \ .
\eeq
Intriguingly, this equation forms a connection between MERW and quantum mechanics, as one may interpret $\vec{\psi}_{0}$ as the wave function of the ground state of the operator $-\mathbf{A}$ and consequently $\psi^2_{0i}$ becomes the probability of finding a particle in this state \cite{ZB1,ZB2}. The two random walks, (\ref{Porw}) and (\ref{Pmerw}), in general exhibit altogether different behaviors except for the case of $k$-regular graphs, where they coincide.

%---------------------------------> SECTION
\section{Cayley tree}
\label{sec:Cayley}

\begin{figure}[bpt!]
	\centering
		\includegraphics[width=0.49\textwidth]{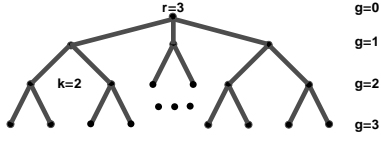}
	\caption{\label{fig:tree} Cayley tree with root degree $r=3$, branching number $k=2$, and $G=3$ generations.}
\end{figure}

We define a Cayley tree with a branching number $k$, which is the number of edges leading from a given node to the next generation of nodes, and the number of generations $G$. The root of the tree is assumed to have a degree $r$ and it belongs to the zeroth generation (see Fig. \ref{fig:tree}). The number of nodes in the zeroth generation is therefore $n_0=1$, in the first $n_1=r$ nodes, in the second $n_2=rk$, in the third one $n_3=rk^2$, etc. The tree has $n$ nodes in total: $n=\sum_{g=0}^G n_g =1+r(k^G-1)/(k-1)$.

%--------------------------------------------------> subSECTION
\subsection{Eigenvalues of the adjacency matrix}
\label{sec:adj}

This section is devoted to calculation of eigenvalues of the adjacency matrix of Cayley tree, which can be determined by solving the equation
\beq
\det(\mathbf{A}-\lambda\mathbf{1})=0
\label{eigen}
\end{equation}
The determinant can be calculated with the use of a sequence of elementary transformations that leave it invariant, \eg, additions of multiple of a row or column to another row or column. Thus, the determinant can be reduced to a triangular form with zeros above the diagonal, as first presented in \cite{Cayley}. Details of this procedure can be found in \cite{ZB3}.
The triangular form of the determinant allows to rewrite \eqref{eigen} as a product of the diagonal coefficients
\beq
\prod_{g=0}^G \left[A_g(\lambda)\right]^{m_g} = 0,
\label{productA}
\eeq
where $m_G=1$ and $m_{G-g} = n_{g} - n_{g-1}$, for $g=1,2,\ldots,G$, and $A_g(\lambda)$ are polynomials w.r.t. $\lambda$ given by the recursive equations
\begin{eqnarray}
A_0(\lambda) & = & -\lambda,  \nonumber \\
A_g(\lambda) & = & -\lambda A_{g-1}(\lambda) - k A_{g-2}(\lambda), \quad \uni{for} \ g<G, \label{polynoms} \\
A_G(\lambda) & = & -\lambda A_{G-1}(\lambda) - r A_{G-2}(\lambda). \nonumber
\end{eqnarray}
Notice that for $g=G$ the coefficient $k$ is replaced by $r$, which is a consequence of the tree structure allowing arbitrary root degree. To complete the set of equations we have to take initial condition $A_{-1}=1$.
The real roots of equation (\ref{productA})
counted with the degeneracy $m_g$ give the total number of $\sum_g (g+1) m_g = \sum_g n_g = n$, which means all $n$ eigenvalues of the adjacency matrix are retrieved.

The recurrence \eqref{polynoms} can be solved:
\beq
A_g = k^{(g+1)/2}\frac{\sin[(g+2)\theta]}{\sin\theta},\ \uni{for} g<G
\label{Ag}
\eeq
where $\cos\theta = -\lambda/(2\sqrt{k})$ and $\theta$ is an auxiliary parameter.
To obtain the polynomial $A_G$ one needs combine the last equation in (\ref{polynoms}) 
$A_G= -\lambda A_{G-1} - k A_{G-2}$ with the solutions for $A_{G-1}$ and $A_{G-2}$ (\ref{Ag}), which yields
\beq
A_G=k^{(G-1)/2} \frac{k \sin [(G+2)\theta]+(k-r) \sin (G\theta)}{\sin\theta}.
\label{AG}
\eeq

Now, instead of \eqref{productA} we can consider equations $A_g=0$ and $A_G=0$ to find the solutions for $\theta$ and then determine the eigenvalues of the adjacency matrix 
using the formula $\lambda = -2\sqrt{k} \cos\theta$.
The first equation, leading to
\beq
\sin[(g+2)\theta]=0,
\eeq
has $g+1$ solutions
\beq
\lambda_{g,j}=2\sqrt{k} \cos\left(\frac{\pi j}{g+2}\right),
\quad \uni{for} \ j=1,\ldots, g+1.
\label{gj}
\eeq
As follows from (\ref{productA}), each eigenvalue $\lambda_{g,j}$ in this series has multiplicity $m_g$. The equation $A_G=0$ produces
\beq
k \sin [(G+2)\theta]+(k-r) \sin (G\theta) =0,
\label{thetaG}
\eeq
for which analytical solutions exist in the case $r=k$ 
\beq
\lambda_{G,j}=2\sqrt{k} \cos\left(\frac{\pi j}{G+2}\right),
\quad \uni{for} \ j=1,\ldots, G+1,
\label{Gj1}
\eeq
and in the case $r=2k$
\beq
\lambda_{G,j}=2\sqrt{k} \cos\left[\frac{\pi (j-1/2)}{G+1}\right],
\quad \uni{for} \  j=1,\ldots, G+1 \ .
\label{Gj2}
\eeq
Other choices of $r$ involve numerical solving of (\ref{thetaG}).

It can be shown that the largest eigenvalue of the adjacency matrix is
$\lambda_0=\lambda_{G,1}$. It belongs to one of three classes of solutions depending on $r$, which takes values $r \in (0,2k - 2k/G)$ in the first class, $r \in (2k-2k/G,2k+2k/G)$ in the second,
and $r \in (2k+2k/G,+\infty)$ in the third.

For large $G$ (\ie, $G \gg 2k$) the second interval becomes just a single integer value $r=2k$. The first class allows values $r < 2k$ for which an approximate solution exists
\beq
\lambda_0 = 2\sqrt{k} \cos \frac{\pi}{G+\delta},
\label{Gplusdelta}
\eeq
where $\delta \approx 2k/(2k-r)$, or the exact solution for $r=k$ \eqref{Gj1}.
In the third class, $r>2k$, there are no real 
solutions of (\ref{thetaG}) for $\theta\in(0,\pi/(G+1))$ and $\lambda_0$ corresponds to a purely imaginary $\theta$. 
The trigonometric equation \eqref{thetaG} is thus replaced by a hyperbolic one.
In the limit of large $G$ the approximate solution is
\beq
\lambda_0 = \frac{2\sqrt{k}}{\sqrt{1-x^2}},
\label{lambdax}
\eeq
where
\beq
x = z \left[ 1 -2\left(\frac{1-z}{1+z}\right)^{G+1}\right] \ \uni{and}\ z= 1-\frac{2k}{r} \ .
\label{x}
\eeq

%------------------------------------> new section:6
\subsection{The eigenvalues of the GRW transition matrix}

The stochastic matrix of generic random walk (\ref{Porw}) can be subjected to the same procedure as explained in Sec. \ref{sec:adj}. Transforming its determinant to the triangular form generates analogous recursion as in \eqref{polynoms}:
\begin{align}
A_0&=-\lambda, \nonumber \\
A_g&=-\lambda A_{g-1}-\frac{k}{(k+1)^2} A_{g-2}, \quad \uni{for} \ g=2,\ldots, G\!-\!1, \\
A_G&=-\lambda A_{G-1}-\frac{1}{k+1}A_{G-2}. \nonumber
\label{eq:rekGRW}
\end{align}
We take $A_{-1}=k+1$ as an initial condition that agrees with the rest of equations and proceed as before solving this recurrence to obtain eigenvalues from the equations $A_g=0$ and $A_G=0$.
From $A_G=0$ one gets
\beq{}
2k\cos(2\theta)=1+k^2 \ \uni{and} \ \sin(G \theta)=0,
\eeq
whose solutions lead to, respectively,
\beq
\lambda_0=1,\ \uni{and}\ \lambda_{G,j}=2\sqrt{ \frac{k}{(k+1)^2}}\cos\left(\frac{\pi j}{G}\right), \quad \uni{for} \ j = 1,\ldots,G,
\eeq
and from $A_g=0$
\beq
	\sin [(g+2)\theta]=k \sin(g\theta).
	\label{thetaGGRW}
\eeq
The last equation has identical form as (\ref{thetaG}) except for different coefficients. The class of solutions of (\ref{thetaG}) with $r \in (2k+2k/G,+\infty)$ corresponds to value $k>1$ in the above equation. Hence, the value of $\theta$ that leads to the largest eigenvalue in a given series is imaginary. Once again, the trigonometric equations \eqref{thetaGGRW} change into hyperbolic ones. Upon replacements $k=\frac{z+1}{1-z},\ z=\frac{k-1}{k+1}$, we end up with (\ref{x}) rewritten as $x=\frac{k-1}{k+1}\left[1-2k^{-(g+1)}\right]$.
In large $G$ limit, the second largest eigenvalue is thus approximated by
\beq
\lambda_1= 2\sqrt{\frac{k}{(k+1)^2}} \frac{1}{\sqrt{1-x^2}}
\label{lambdaxGRW}
\eeq
and clearly the second largest eigenvalue $\lambda_1$ approaches $\lambda_0=1$ exponentially in $G$.

%------------------------------------> new section:4
\subsection{Stationary states of GRW and MERW on Cayley trees}

A random walk on a graph has a stationary probability distribution if the graph is not bipartite. If a graph is bipartite, one can define a semi-stationary state: it involves either averaging probability distributions over two consecutive time steps $t$ and $t+1$ (because the distributions for even and odd times are independent) or averaging the distribution over initial conditions.

GRW leads to the stationary occupation probabilities 
\beq
\pi_i = \frac{k_i}{\sum_j k_j}, \quad \uni{for} \ i=1,\ldots,n,
\eeq
which comprise a flat distribution for nodes of degree $k_i=k+1$ and the exception of the root having $r$ neighbors and leaves neighboring with just one node. As nodes in each generation have equal stationary probabilities we can sum over them $\Pi_g =n_g \pi_{i} \propto k^{g-1}$, which produces the exponential factor.

The stationary probabilities of MERW are equal to the squared components of $\vec{\psi}_0$. All elements $\psi_{0i}$ of this vector have the same values for $i$ belonging to a given generation $g$. This simplifies the description of the stationary state so that we may write $\psi_g$ for all nodes in the generation $g$ (we omit the first index, which numbers the corresponding eigenvalue). Exact solution for $\psi_g$ can be obtained by solving a recurrence equation analogous to \eqref{polynoms}:
\beq
\pi_i = \psi^2_{0i} \propto k^{G-g} \sin[(G-g+1)\theta]^2, \quad \uni{for} \ g=0,\ldots,G \ \uni{and} \ i \in g,
\eeq
where the normalization constant has been omitted. After summing over whole generation $i\in g$, the probabilities become
\beq
\Pi_{g}=n_g \pi_{i}\propto k^{G-1} \sin[(G-g+1)\theta]^2, \quad \uni{for} \ g=1,\ldots,G \ \uni{and} \ i \in g,
\eeq
where the case $g=0$ with its $n_0=1$ needs a separate treatment. 

This result depends on the choice of $r,k$ through $\theta$ and $\lambda$. For $r<2k$, parameter $\theta \approx \frac{\pi}{G+\delta}$ and the limiting distribution is a sine square; for $r=2k$, $\theta = \frac{\pi/2}{G+1}$ and the distribution is a cosine square; for $r>2k$, $\theta = i\ \uni{arctanh}\ x$ [where $x$ is defined in (\ref{x}), while $i$ is the imaginary unit], which yields a hyperbolic sine. 
These limiting results as well as finite-size effects are showed in an online interactive demonstration \cite{Demo1}.

%------------------------------------> new section:5

\subsection{Relaxation times}

A stochastic matrix needs not be symmetric, thus its
right and left eigenvectors may differ: $\mathbf{P}\vec{\Psi}_\alpha = \Lambda_\alpha \vec{\Psi}_\alpha,
\vec{\Phi}_\alpha \mathbf{P} = \Lambda_\alpha \vec{\Phi}_\alpha$.
Hence, there exists a spectral decomposition of $\mathbf{P}$
\beq
P_{ij} = \sum_\alpha \Lambda_\alpha \Psi_{\alpha i} \Phi_{\alpha j} \ ,
\label{sd}
\eeq
where for MERW one can make replacements: $\Lambda_\alpha = \lambda_\alpha/\lambda_0, \Psi_{\alpha i} = \psi_{\alpha i}/\psi_{0 i}$, $\Phi_{\alpha i} = \psi_{\alpha i} \psi_{0 i}$.
The spectral decomposition of the adjacency matrix of a given graph thus contains information about all properties of MERW.

From the knowledge of the initial probability distribution $\vec{\pi}(0)$ and the transition matrix $\mathbf{P}$ the distribution  can be determined at any time $t$
\beq
\vec{\pi}(t) = \vec{\pi}(0) \mathbf{P}^t ,
\eeq
where the elements $\pi_i(t),\ i=1,\ldots,n$ of $\vec{\pi}(t)$ denote the probability of finding a particle performing a random walk at a node $i$ at time $t$.

The last equation can be reformulated utilizing the spectral decomposition of the stochastic matrix (\ref{sd})
\beq
\vec{\pi}(t) = \sum_\alpha c_\alpha \Lambda_\alpha^t \vec{\Phi}_\alpha \ .
\label{sd2}
\eeq
where $c_\alpha$ denotes a spectral coefficient: $c_\alpha=\vec{\pi}(0)\cdot \vec{\Psi}_\alpha = \sum_i \pi_i(0) \Psi_{\alpha i}$.

Generally, all eigenvalues $\Lambda_\alpha$ of $\mathbf{P}$
are located inside or on the unit circle in the complex plane $|\Lambda_\alpha|\le 1$ and in the limit of infinite $t$ on the right-hand side of \eqref{sd2} only $|\Lambda_\alpha|=1$ survive, while all the other terms vanish exponentially.

For both GRW and MERW on a tree only two eigenvalues on the unit circle are left
$\Lambda_0=1$ and $\Lambda_n=-1$ due to bipartiteness of the graph. For $t \rightarrow \infty$ the relaxation to the stationary state is generically governed by the second largest eigenvalue
$\Lambda_1$ and its negative counterpart $\Lambda_{n-1}=-\Lambda_1$.
The corresponding term in the spectral decomposition (\ref{sd2})
decreases exponentially as $\exp(-t/\tau_1)$, where $\tau_1=[-\ln(\Lambda_1)]^{-1} = \left[\, \ln(\lambda_0/\lambda_1)\right]^{-1}$. 

Thus, $\tau_1$ is what we call the generic relaxation time, which is the largest one. We note, however, that there are symmetries that lead also to other relaxation times.
As the eigenvalues of the adjacency matrix depend on the tree parameters, also the relaxation times for MERW fall into several classes. The relaxation times for large $G$ are given in the Table \ref{tab:relax}. It is noteworthy that whereas the probability distribution for GRW relaxes linearly with the system size $\tau_1\sim n\sim k^G$, for MERW it is as fast as a logarithm of the system size $\tau_{1}\sim \ln n$.
Derivations and further details expanding the note on symmetries can be found in \cite{ZB3}. An online interactive demonstration \cite{Demo2} can also facilitate understanding of these results.

\begin{table}%
\begin{tabular}{l|lll}
Regime & $\lambda_0$ & $\lambda_1$ & $\tau_1$ \\
\hline
Strongly branched:\\ $r>2k,\ k>1$ & Eq.\eqref{lambdax}&$\lambda_{G-1,1}$& $c+\frac{c^2 \pi^2}{2}\frac{1}{G^2}+\ldots$\\
Critically branched:\\ $r=2k,\ k>1$& $\lambda_{G,1}$&$\lambda_{G-1,1}$&$\frac{8 G^2}{3 \pi^2}+\frac{16 G}{3\pi^2}+\ldots$\\
Weakly branched:\\ $1<r<2k,\ k>1$&$\lambda_{G,1}$&$\lambda_{G-1,1}$& $\frac{2k-r}{r \pi^2}G^3+\frac{3(4k-r)}{2 r \pi^2}G^2+\ldots$\\
Planted tree:\\ $r=1$ &$\approx$ Eq.\eqref{Gplusdelta}&$\lambda_{G-2,1}$&$\frac{2k-1}{2k\pi^2}G^3+ \frac{3}{2\pi^2}G^2+\ldots$\\
Linear chain:\\ $k=1,\ r=1$&$\lambda_{G,1}$&$\lambda_{G,2}$&$\frac{2 G^2}{3\pi^2}+\frac{8 G}{3\pi^2}+\ldots$\\
GRW: $r>1,\ k>1$ &1& Eq.(\ref{lambdaxGRW})&$\frac{8k}{(k-1)^2} k^{G}$\\

\end{tabular}
\caption{\label{tab:relax} Relaxation times $\tau_1$ for large $G$. All rows except for the last one refer to MERW. The symbols $\lambda_{g,j}$ correspond to one of the equations \eqref{gj},\eqref{Gj1}, or \eqref{Gj2}, whichever is appropriate for the choice of parameters $k,r$. In the first row: $c=(\ln \frac{r}{2\sqrt{(r-k)k}})^{-1}$. While the number of vertices $n\sim k^G$, the probability distribution relaxes a logarithm of the system size $\tau_{1}\sim \ln n$.}
\end{table}

\section{Ladder graph}

\begin{figure}[bpt!]
\centering
\includegraphics[width=1.\textwidth]{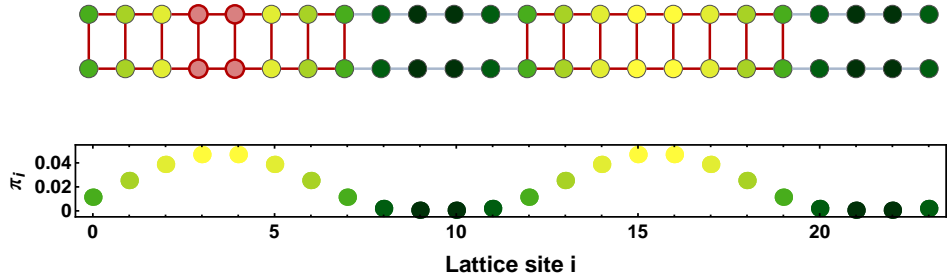}
\caption{\label{fig:drabinka} (Colors online) Stationary probability for a ladder graph with periodic boundary conditions: the probability localizes in the intact regions. Red nodes represent initial condition that would be chosen for this graph.}
\end{figure}

In this section, we discuss a particular class of ladder graphs (exemplary ladder graph can be seen in Fig.\ref{fig:drabinka}). A ladder graph consists of two chains of integer length $n/2$ which are connected by rungs, \ie node $i$ of one chain is connected to node $i'$ of the second one, then $i+1$ to $i'+1$ and so forth. We also impose periodic boundary conditions producing a ring, where node $i+n/2$ is connected to node $i+1$, and node $i'+n/2$ to node $i'+1$. This structure is symmetric with respect to reflection $i \rightarrow i'$, and so the graph is a quasi one-dimensional system. It is a 3-regular graph, although we remove some rungs from the ladder to introduce defects, so that MERW and GRW are not equivalent on this graph anymore.

For the adjacency matrix of the graph $\mathbf{A}$, its largest eigenvalue $\lambda_0$ and the eigenvector $\vec{\psi}_0$ associated with it, the stationary solution for MERW is given as the ground state of the tight-binding equation
\beq
\left(\mathbf{H} \vec{\psi}_0\right)_a=\left(-\Delta \vec{\psi}_0\right)_a+V_a\psi _{0,a}=E_0\psi _{0,a}
\eeq
where the Hamiltonian is defined as $H_{ab}=k_{\uni{max}}\delta_{ab}-A_{ab}$, with the Kronecker delta $\delta_{ab}$, maximum degree of the graph $k_{\uni{max}}$, and $V_a=k_{\max }-k_a,\, E_0=k_{\max }-\lambda _0$.
For a ladder graph with defects this equation yields
\beq
2\psi _{0,a}-\psi _{0,a-1}-\psi _{0,a+1}+V_a\psi _{0,a}=E_0\psi _{0,a}
\eeq
where $E_0=3-\lambda_{0}$ and $V_a=0$ or $1$ (rung present or absent).
Stationary states of a number of ladder graphs (with one, two, or a number of random defects) were discussed in Section 6 of \cite{ZB2}.

Additionally, we impose a symmetry on those defects: there can only be two equal regions intact and two equal regions with rungs removed (gaps). We take the initial probability $1$ at the center of one of the intact regions (this may be $2$ or $4$ nodes, depending on whether the length of the region is odd or even, see Fig.\ref{fig:drabinka}). The systems under study have $n= 48-512$ total number of nodes and the number of deleted rungs separating two regions (the gap size) varies between $g=1-10$.

We measure the probability $P(t)$ summed over one whole region (as the regions are equal in size, $P_{\infty}=1/2$ is its stationary value). It might be understood as a macroscopic measure of the process taking place in this region. As expected, the probability flows from one the initial intact region to the other one until equilibrium ($P(t)=1/2$ in both regions) is attained.

\begin{figure}[bpt!]
	\centering
\includegraphics[width=0.49\textwidth]{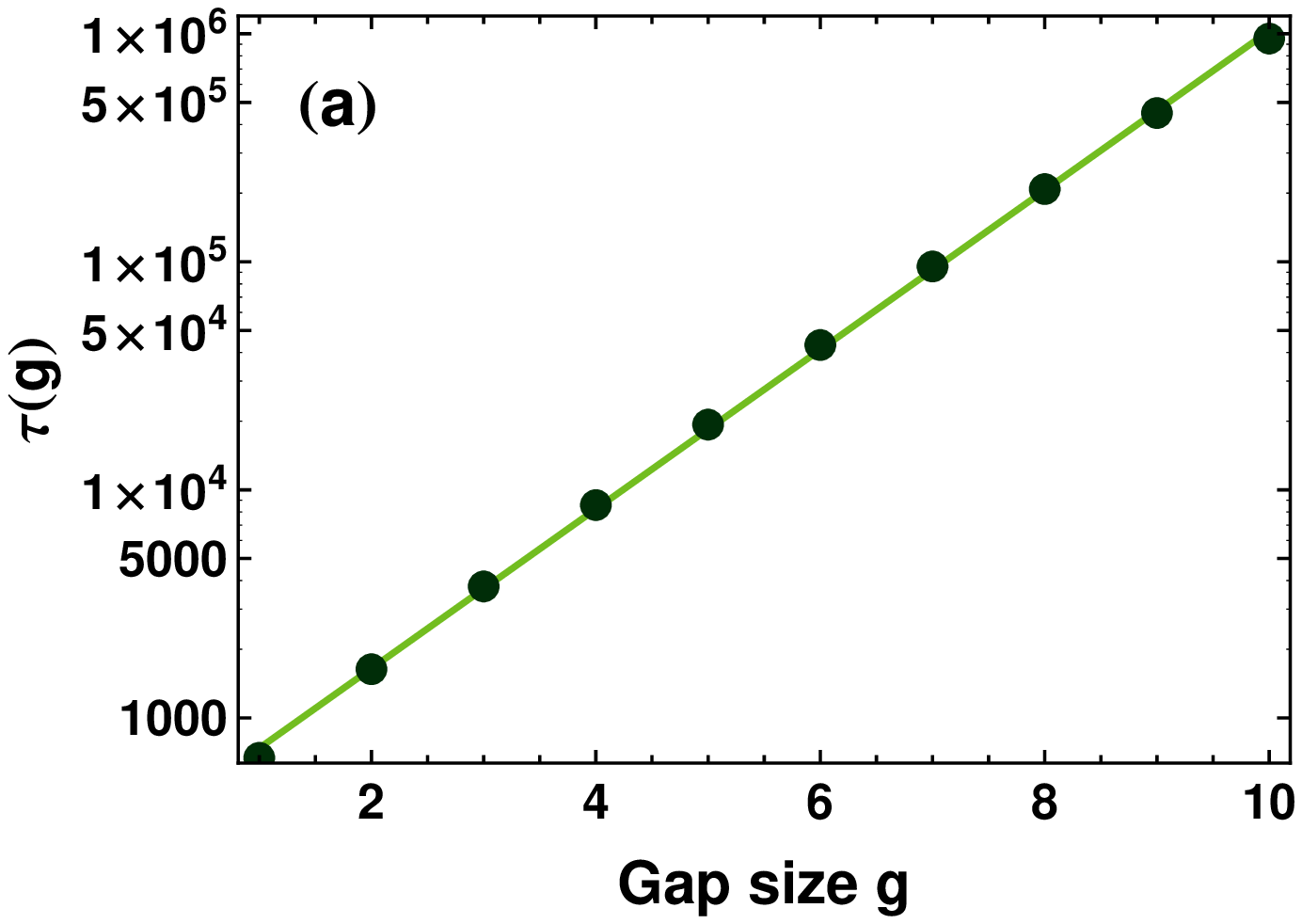}
\hfill
\includegraphics[width=0.49\textwidth]{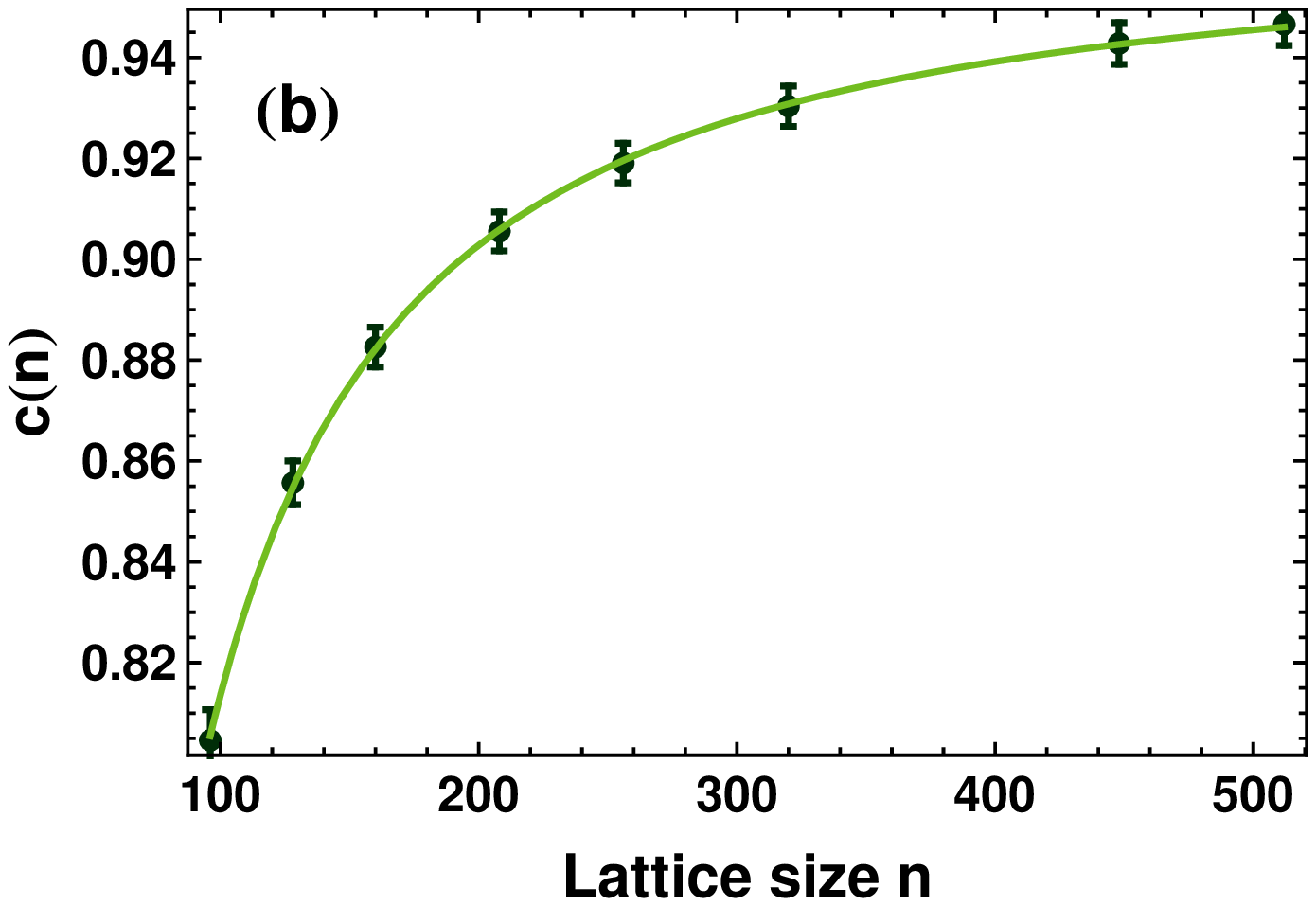}
	\caption{\label{fig:LadderFitMERW} Maximal Entropy Random Walk: (a) Logarithmic plot shows an exponential dependence of the relaxation time on the gap size (an exemplary system size, $n=96$) reminding of quantum tunneling, (b) the dependence of the relaxation time on the system size, $c(n)=c_{\infty}-f n ^{-1/\nu }$ [see \eqref{eq:ladder}]. Continuous lines are the best fits of an exponential function and power law, respectively.}
\end{figure}

We fit the numerical results to exponential dependence on time $t$: $P(t)\sim \exp [-a(t-b)]$, where $a$ and $b$ are fitting parameters from which we extract the relaxation time $\tau$, which is the characteristic time scale of an exponential approach to the stationary state.
The results for the behavior of relaxations times for GRW and MERW are given in Table \ref{tabela}.
It turns out that for MERW there is a clear dependence of the relaxation on the gap size for a given lattice size (example in Fig. \ref{fig:LadderFitMERW}(a) for $n=96$): $a(g)=\exp (-c\cdot g-d)$, $b(g)=\exp (c\cdot g+d)$, where $c$, $d$ are constants with respect to the gap size $g$. After extracting this dependence, only the dependence on the system size remains in the function $c=c(n)$, which is very well fitted with a power law [Fig. \ref{fig:LadderFitMERW}(b)]:
$c(n)=c_{\infty}-f n ^{-1/\nu }$ (best-fit value parameters are $c_{\infty}=0.9643\pm0.0078$, $f=58\pm42$, $\nu=0.773\pm 0.098$). Thus, the macroscopic probability depends on time, system size, and gap size:
\beq
|P(t;g,n)-P_{\infty}| \propto \exp \left\{-\exp \left[-c(n)\cdot g\right]\cdot t\right\}.
\label{eq:ladder}
\eeq

\begin{figure}[bpt!]
	\centering
\includegraphics[width=0.495\textwidth]{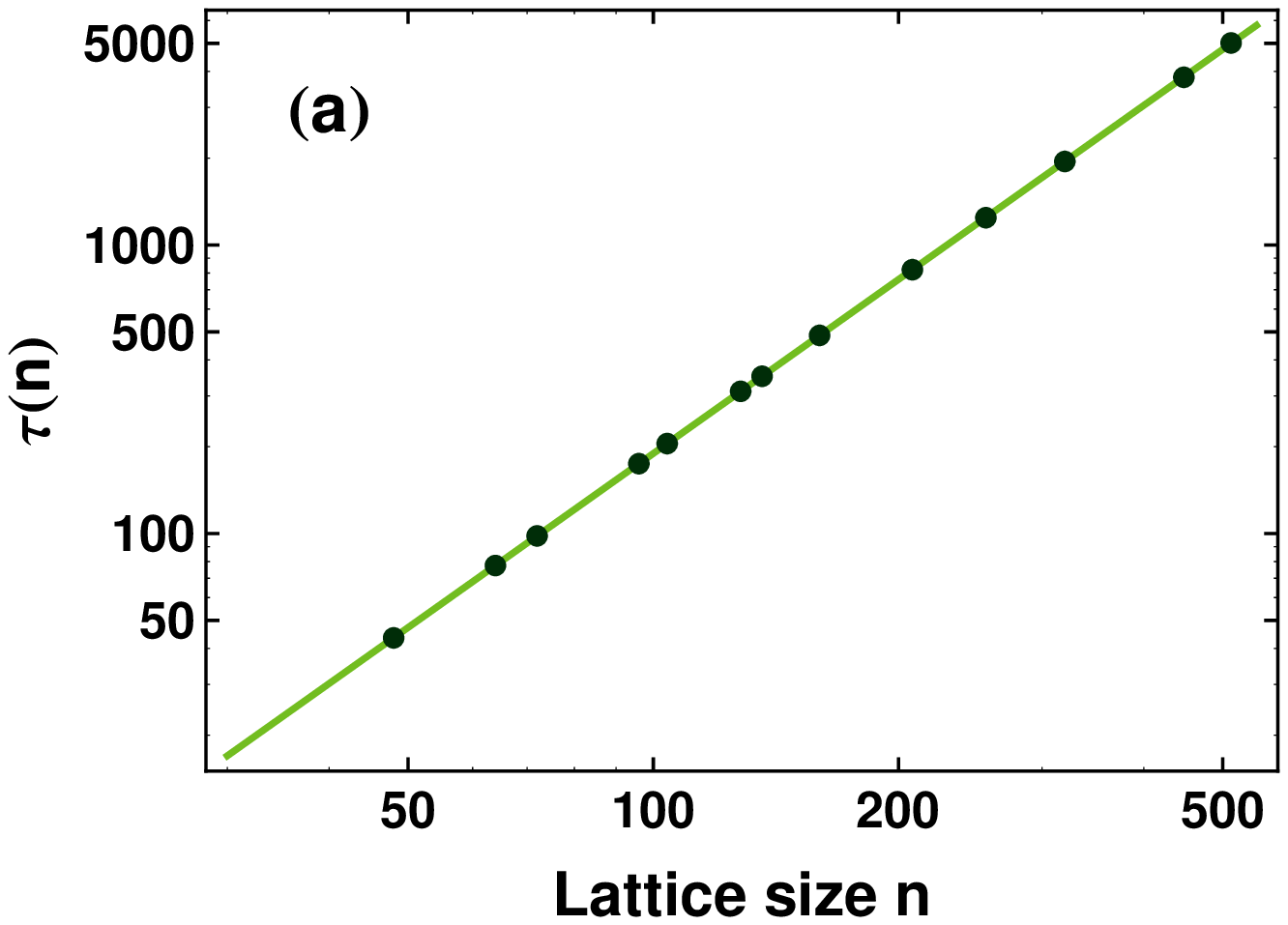}
\hfill
\includegraphics[width=0.485\textwidth]{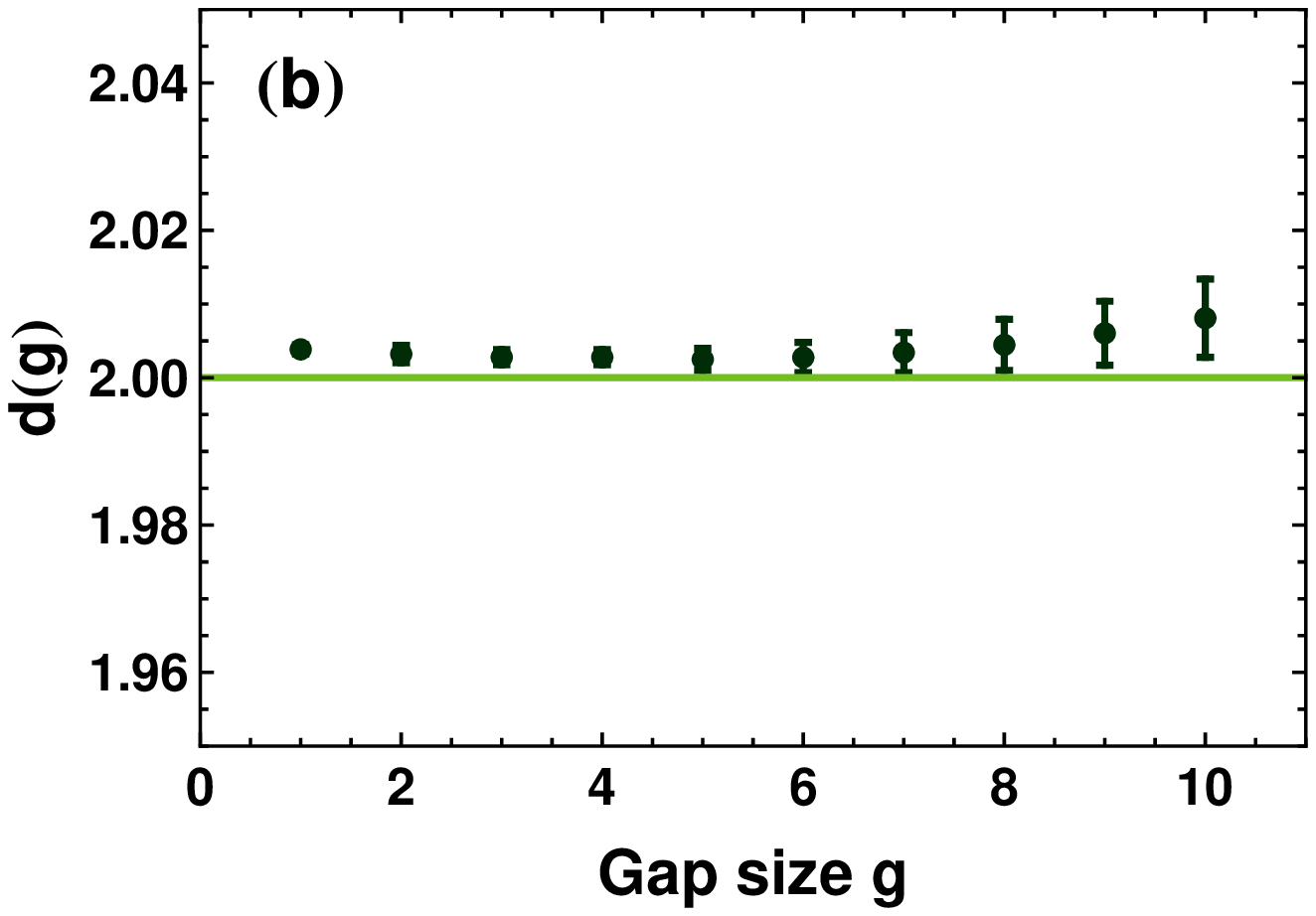}
	\caption{\label{fig:LadderFitGRW} Generic Random Walk:(a) log-log plot shows power law dependence of relaxation time, expected for a diffusion process (continuous line is the best fit; gap size $g=1$), (b) the best-fit exponents $d$ of the power law show independence from the gap size. The errors result from finite-size effects.}
\end{figure}

For GRW, both fitted parameters $a$ and $b$ have shown no dependence from the gap size $g$, although they do depend on the system size: $a(n)=c n^{-d}$, $b(n)=c' n^{d'}$, where $d', d\approx 2$ (see Fig.\ref{fig:LadderFitGRW}). This produces the familiar behavior $\tau \sim n^2$ which is expected for a one-dimensional random walk.

\begin{table}
\begin{center}
\begin{tabular}{c|l}
GRW &  $ \tau(n,g)=c\cdot n^{d},\quad d=\text{const}.=2$\\
MERW & $\tau(n,g) =\exp \left[c(n)\cdot g\right]$
\end{tabular}
\end{center}
\caption{\label{tabela}} Relaxation times $\tau$ as functions of the system size $n$ and gap size $g$, where $c(n)=c_{\infty}-f n ^{-1/\nu }$ and $d, c_{\infty}, f, \nu$ are fitted constant.
\end{table}

\section{Conclusions}

In this paper, we have discussed the dynamics of generic random walk and maximal entropy random walk on two classes of graphs.
For Cayley trees, we have shown the analytic form of generic relaxation times governing how fast probability distributions of those random walks approach their stationary states. MERW has proven to be generically faster (logarithmic with respect to the system size) than GRW (linear w.r. to the system size).
However, on defective ladder graphs the relaxation of probability seems to show opposite behavior: while GRW relaxes diffusively, the relaxation times for MERW are much longer, growing exponentially with the size of the defective region. 

These results indicate that MERW might exhibit comparatively fast relaxation within intact or homogeneous regions (like a Cayley tree) but inhibits the relaxation process between regions separated by defects, bottlenecks or bridges.
While qualities of MERW's stationary states have already been utilized to improve centrality measures in complex networks \cite{MERW+CN2}, its dynamic properties and a close relation between eigenvalues of the adjacency matrix and the statistics of paths may be of use in community search algorithms on complex networks (a number of algorithms based on random walks, path enumeration and spectral properties of the adjacency matrix are reviewed in \cite{F}).
As a more speculative idea, it is also worth remembering that MERW keeps all paths of a given length between any two endpoints equiprobable, which makes it capable of hiding the route information travels, \eg over the Internet. 

%------------------------------------> new SECTION
\section*{Acknowledgments}
The author would like to thank Z. Burda and B. Waclaw for fruitful discussions.
Project operated within the Foundation for Polish Science International Ph.D. Projects Programme co-financed by the European Regional Development Fund covering, under the agreement no. MPD/2009/6, the Jagiellonian University International Ph.D. Studies in Physics of Complex Systems.

\bibliographystyle{plain}

\end{document}